\documentclass[twocolumn,english,pre]{revtex4-1}
\usepackage[T1]{fontenc}
\usepackage[latin9]{inputenc}
\usepackage{amsmath}
\usepackage{graphicx}
\usepackage{esint}

\makeatletter

\providecommand{\tabularnewline}{\\}

\makeatother

\usepackage{babel}
\begin{document}

\title{Surface Forces on a Deforming Ellipsoid in Shear Flow}

\author{E.P. Kightley\textsuperscript{1,2}, A. Pearson\textsuperscript{1,2},
J. A. Evans\textsuperscript{3}, and D.M. Bortz\textsuperscript{1}}
\email{dmbortz@colorado.edu}

\affiliation{1. Department of Applied Mathematics, University of Colorado Boulder~\\
2. Interdisciplinary Quantitative Biology Graduate Program, University
of Colorado Boulder~\\
3. Department of Aerospace Engineering, University of Colorado Boulder}
\begin{abstract}
We present a model for computing the surface force density on a fluid
ellipsoid in simple shear flow, which we derive by coupling existing
models for the shape of a fluid droplet and the surface force density
on a solid ellipsoid. The primary contribution of this coupling is
to develop a method to compute the force acting against a plane intersecting
the ellipsoid, which we call the \emph{fragmentation force}. The model
can be used to simulate the motion, shape, surface force density,
and breakage of fluid droplets and colloidal aggregates in shear flow.
\end{abstract}
\maketitle

\section{Introduction}

Emulsions, in which one fluid (the \emph{dispersion phase}) is dispersed
in another (the \emph{continuous phase}), are central to many industrial
applications, such as the production of polymer blends \cite{Cooper2000JMaterChem},
dielectric materials \cite{JiangBertoneHwangEtAl1999ChemMat}, and
drug manufacturing and delivery \cite{SoppimathAminabhaviKulkarniEtAl2001JControlRelease,MullerMaderGohla2000EurJPharmBiopharm}.
In these applications it is often desirable to describe the behavior
of the dispersion phase as a function of the physical properties and
flow regime of the continuous phase. In particular, significant efforts
have been devoted to modeling the breakage and resulting size distributions
of dispersed droplets (see \cite{SolsvikTangenJakobsen2013RevChemEng}
for a review). These phenomena can depend upon the motion and shape
of the dispersed droplets, topics which have therefore also received
considerable attention (see \cite{Minale2010RheolActaa} for a review). 

Many of these models treat the dispersed particles as ellipsoidal
droplets and perform analyses of motion and breakage on individual
ellipsoids. Such models are designed primarily to simulate two-fluid
emulsions, such as oil in water, but there exist many industrial applications
in which the dispersion phase is itself an inhomogenous colloid. A
major class of such applications involves microbial aggregates in
suspension \cite{Bratby2008,LissDroppoLeppardEtAl2007,Nopens2005,BortzJacksonTaylorEtAl2008BullMathBiol}.
These aggregates may be treated like ellipsoidal droplets for the
purposes of approximating their shape and motion \cite{Blaser2000JColloidInterfaceSci,JamesYogachandranLoewenEtAl2003JPulpPapSci,Blaser2000ColloidsSurfPhysicochemEngAsp,Kobayashi2005WaterRes,Kobayashi2004ColloidsSurfPhysicochemEngAsp},
but it is not clear that the corresponding breakage and size-distribution
models from the rheology literature on emulsions are equally applicable.
The inhomogenous nature of such aggregates may mean that some breakage
patterns are more likely than others, and we may wish to use knowledge
about the structure and composition of the colloid when modeling how
and where they break. 

To this end, we present a model to compute the force acting to break
an ellipsoidal droplet at a specified location, which we call a \emph{fragmentation
force}. This is an extension of our earlier work in which we began
to develop a framework for identifying likely breakage locations in
suspended microbial aggregates \cite{ByrneDzulSolomonEtAl2011PhysRevE}.
The present work expands upon this by introducing deformation to the
model and by refining the computation of the fragmentation force.
We construct our model by coupling a model for the deformation of
a fluid droplet \cite{JacksonTuckerIII2003JRheol,WetzelTuckerIII2001JFluidMech}
(hereafter, the \emph{Deformation Constituent Model }or DCM) with
one that computes the surface force density on a solid ellipsoid \cite{Blaser2002ChemEngSci}
(hereafter, the \emph{Force Constituent Model }or FCM). We restrict
ourselves to the case of viscous shear under the assumption of Stokes'
flow, and our choice of deformation model is further guided by the
requirements that (1) the shape of the droplet remain ellipsoidal
and (2) there be a restorative force (in this case interfacial tension)
acting to oppose the deformation imposed by the shear field.

In Section \ref{sec:Model} we describe the deformation (\ref{subsec:DFM})
and force (\ref{subsec:FCM}) constituent models, discuss the scheme
by which we couple them (\ref{subsec:connect}), define the fragmentation
force (\ref{subsec:connect}), and introduce the parameters of the
model (\ref{subsec:parameters}). In Section \ref{sec:Results} we
present simulations of the model, first discussing some characteristic
behaviors of the DCM and comparing the motion of a deforming ellipsoid
to that of a solid ellipsoid as prescribed in the FCM (\ref{subsec:resultsdef}),
and then examining the behavior of the surface force density and the
fragmentation force (\ref{subsec:resultsfragforce}). Finally, in
Section \ref{sec:Conclusion}, we conclude with a discussion of the
future application of this model to our work on microbial fragmentation.

\section{Model\label{sec:Model}}

\subsection{Deformation constituent model\label{subsec:DFM}}

The DCM is the model we use to describe the deformation and rotation
motion of a fluid ellipsoid in a flow. This model is developed in
\cite{WetzelTuckerIII2001JFluidMech,JacksonTuckerIII2003JRheol},
and in the remainder of this section we summarize the work therein.
An arbitrary ellipsoid centered at the origin can be represented by
a \emph{shape tensor}, a symmetric $3\times3$ tensor $\mathbf{G}$
such that $\mathbf{x}^{T}\mathbf{G}\mathbf{x}=1$ for any point $\mathbf{x}$
on the surface of the ellipsoid. The shape tensor is orthogonally
diagonalizable, so that 
\begin{equation}
\mathbf{D}=\mathbf{R}^{T}\mathbf{G}\mathbf{R}\label{eq:rotationdef}
\end{equation}
where $\mathbf{D}$ is diagonal and $\mathbf{R}$ is a rotation. We
can choose to construct $\mathbf{G}$ such that the diagonal entries
of $\mathbf{D}$ (the eigenvalues of $\mathbf{G}$) $\lambda_{i}$
are defined by $\lambda_{i}=1/a_{i}^{2}$ where $\mathbf{a}=(a_{1},a_{2},a_{3})$
are the axes lengths of the ellipsoid.

The deformation constituent model consists of an ODE we can solve
for such a shape tensor $\mathbf{G}(t)$. Assuming constant volume
and Stokes flow in an incompressible Newtonian fluid the governing
equation for shape of the ellipsoid is:

\begin{equation}
\mathbf{\dot{G}}+\mathbf{L}_{d}^{\text{T}}\cdot\mathbf{G}+\mathbf{G}\cdot\mathbf{L}_{d}=0\label{eq:gov}
\end{equation}
where $\mathbf{G}$ is the shape tensor of the ellipsoid, as described
in the above, $\mathbf{\dot{G}}$ is the material derivative of $\mathbf{G}$,
and $\mathbf{L}_{d}$ is the velocity gradient inside the droplet. 

In order to solve equation (\ref{eq:gov}) for $\mathbf{G}$, an expression
for $\mathbf{L}_{d}$ is required; this expression must depend only
on the external velocity gradient $\mathbf{L}$, the shape and orientation
of the ellipsoid, and the input parameters. Because of the assumption
of Stokes' flow, the Navier-Stokes equations are linear and so a solution
may be obtained by a superposition of solutions to the separate problems
of (i) a droplet retracting in a vacuum and of (ii) a droplet deforming
in the absence of interfacial tension. The reader is referred to \cite{WetzelTuckerIII2001JFluidMech,JacksonTuckerIII2003JRheol}
for derivations and the precise form of $\mathbf{L}_{d}$.

\subsection{Force constituent model\label{subsec:FCM}}

Here we summarize the FCM applied to simple shear flow, which is developed
in \cite{Blaser2002ChemEngSci}. Given an ellipsoid with axes lengths
$a_{i}$, such that $a_{1}\geq a_{2}\geq a_{3}$, under the assumption
of Stokes' flow, the force density on the surface of a solid ellipsoid
in simple shear can be written as

\begin{equation}
\mathbf{f}=\left(-p_{0}\mathbf{I}-4\mu\sum_{j=0}^{3}\chi_{j}A_{j}^{j}\mathbf{I}+\frac{8\mu}{a_{1}a_{2}a_{3}}\mathbf{A}^{T}\right)\mathbf{n}\label{eq:force}
\end{equation}
where $p_{0}$ is pressure, $\mu$ is the matrix viscosity, $a_{i}$
are the axes lengths, and $\mathbf{n}$ is normal to the surface of
the ellipsoid. The matrix $\mathbf{A}$ in equation (\ref{eq:force})
is defined by 

\begin{equation}
A_{k}^{i}=\begin{cases}
\frac{3\chi_{i}''E_{i}^{i}-\sum_{l=1}^{3}\chi_{l}''E_{l}^{l}}{6(\chi_{1}''\chi_{2}''+\chi_{1}''\chi_{3}''+\chi_{2}''\chi_{3}'')} & \text{ for }i=k,\\
\frac{\chi_{i}E_{k}^{i}+a_{k}^{2}\sum_{l=1}^{3}\varepsilon^{ikl}\chi_{l}'(\varepsilon^{ikl}\Omega_{k}^{i}+\omega_{l})}{2(a_{k}^{2}\chi_{k}+a_{i}^{2}\chi_{i})\sum_{l=1}^{3}|\varepsilon^{ikl}|\chi_{l}'} & \text{ for }i\ne k
\end{cases}\label{eq:A}
\end{equation}
where $\mathbf{E}=\frac{1}{2}(\mathbf{L}+\mathbf{L}^{\text{T}})$
is the rate-of-strain tensor, $\mathbf{\boldsymbol{\Omega}}=\frac{1}{2}(\mathbf{L}-\mathbf{L}^{\text{T}})$
is the vorticity tensor, and $\omega_{l}$ is the $l$th component
of the angular velocity $\boldsymbol{\omega}$ of the ellipsoid. The
elliptic integrals $\chi_{j}$ used in equation (\ref{eq:A}) are
defined by 
\begin{equation}
\chi_{j}=\intop_{0}^{\infty}\frac{d\xi}{(a_{j}^{2}+\xi)\sqrt{(a_{1}^{2}+\xi)(a_{2}^{2}+\xi)(a_{3}^{2}+\xi)}}\label{eq:ellip1}
\end{equation}
with 

\begin{align}
\chi_{i}^{'}=\frac{\sum_{k,l=1}^{3}\varepsilon^{ikl}(\chi_{l}-\chi_{k})}{\sum_{k,l=1}^{3}\varepsilon^{ikl}(a_{k}^{2}-a_{l}^{2})}\label{eq:ellip2}\\
\chi_{i}''=\frac{\sum_{k,l=1}^{3}\varepsilon^{ikl}(a_{k}^{2}\chi_{k}-a_{l}^{2}\chi_{l})}{\sum_{k,l=1}^{3}\varepsilon^{ikl}(a_{k}^{2}-a_{l}^{2})}\label{eq:ellip3}
\end{align}

\subsection{Coupling the models\label{subsec:connect}}

The matrix $\mathbf{A}$ depends upon the matrix velocity gradient
$\mathbf{L}$ and the angular velocity $\boldsymbol{\omega}$ of the
ellipsoid, both of which must be expressed in a frame of reference
relative to the center of the ellipsoid; i.e., one that rotates with
respect to the external frame of reference. In the case of a solid
ellipsoid in shear flow, there are analytic representations for both
of these quantities \cite{Blaser2002ChemEngSci}, but in our model,
the motion of the ellipsoid is dictated by the deformation constituent
model, and we must therefore compute $\mathbf{L}$ and $\boldsymbol{\omega}$
numerically as they do not have closed-form solutions. The rotation
connecting the two reference frames is represented by the matrix $\mathbf{R}(t)$
in equation (\ref{eq:rotationdef}) which we obtain by diagonalizing
the solution $\mathbf{G}(t)$ to the DCM, equation (\ref{eq:gov}).
In simple shear flow, the velocity gradient $\mathbf{L}$ is constant
in time in an external frame of reference. If the ellipsoid rotates
according to $\mathbf{R}(t)$ in the external frame, then the shear
field rotates according to $\mathbf{R}^{T}(t)$ in the ellipsoid frame.
Thus we set $\mathbf{L}_{R}(t)=\mathbf{R}(t)\mathbf{L}\mathbf{R}^{T}(t)$,
and use $\mathbf{L}_{R}$ in equation \ref{eq:force}. Writing the
angular velocity in the anti-symmetric matrix 
\[
[\boldsymbol{\omega}(t)]_{\times}\equiv\left(\begin{array}{ccc}
0 & -\omega_{z} & \omega_{y}\\
\omega_{z} & 0 & -\omega_{x}\\
-\omega_{y} & \omega_{x} & 0
\end{array}\right)
\]
a straightforward calculation tracking the motion of an arbitrary
point on the ellipsoid surface yields the relation

\begin{equation}
[\boldsymbol{\omega}(t)]_{\times}=\left(\mathbf{R}(t)\mathbf{R}'(t)\right)^{T}.\label{eq:wmat}
\end{equation}
We approximate $\mathbf{R}'(t)$ using the discretized solution $\mathbf{R}(t_{i})$
to equation (\ref{eq:gov}) and then use equation (\ref{eq:wmat})
to compute $[\boldsymbol{\omega}(t)]_{\times}$, giving $\boldsymbol{\omega}$,
the angular velocity of the ellipsoid in the external frame. In the
ellipsoid frame, the shear field is rotating in the opposite direction,
with angular velocity $-\boldsymbol{\omega}$. This is the quantity
we use in equation (\ref{eq:force}). 

\subsection{Fragmentation Force\label{subsec:fragforce}}

We want a way to check for breakage in a fluid ellipsoid given some
additional information about where we expect the ellipsoid might be
more likely to break. To do this we expand upon our previous work
in which we check for breakage along the intersection of a plane with
the ellipsoid \cite{ByrneDzulSolomonEtAl2011PhysRevE}. In practice,
the location and orientation of the plane is to be chosen based upon
structural information about the colloidal ellipsoid. In our aforementioned
application, we preferentially chose planes corresponding to locations
where we expected the surface of the microbial aggregate to exhibit
a more negative Gaussian curvature.

Suppose that we have chosen some plane $P$ defined by defined by
$\mathbf{n}_{p}\cdot(\mathbf{x}_{p}-\mathbf{x})=0$ where $\mathbf{n}_{p}$,
$\mathbf{x}_{p}$ are a normal vector and interior point to $P$.
Let $\mathbf{f}(\mathbf{x})$ be the force density at point $\mathbf{x}$
on the surface $\mathcal{E}(t)$ to an ellipsoid at time $t$, both
computed as described the preceding section. The \emph{fragmentation
force} is defined as
\begin{equation}
F=\int_{\mathcal{E}(t)}s(\mathbf{x},P)\,|\mathbf{f}(\mathbf{x})\cdot\mathbf{n}_{p}|\,d\mathbf{x}\label{eq:fragforce}
\end{equation}

where
\begin{equation}
s(\mathbf{x},P)=\begin{cases}
1 & \text{if }\mathbf{f}(\mathbf{x})\text{ acts to pull against }P\\
-1 & \text{if }\mathbf{f}(\mathbf{x})\text{ acts to push into }P
\end{cases}
\end{equation}

\begin{figure*}
\centering\includegraphics[width=16cm]{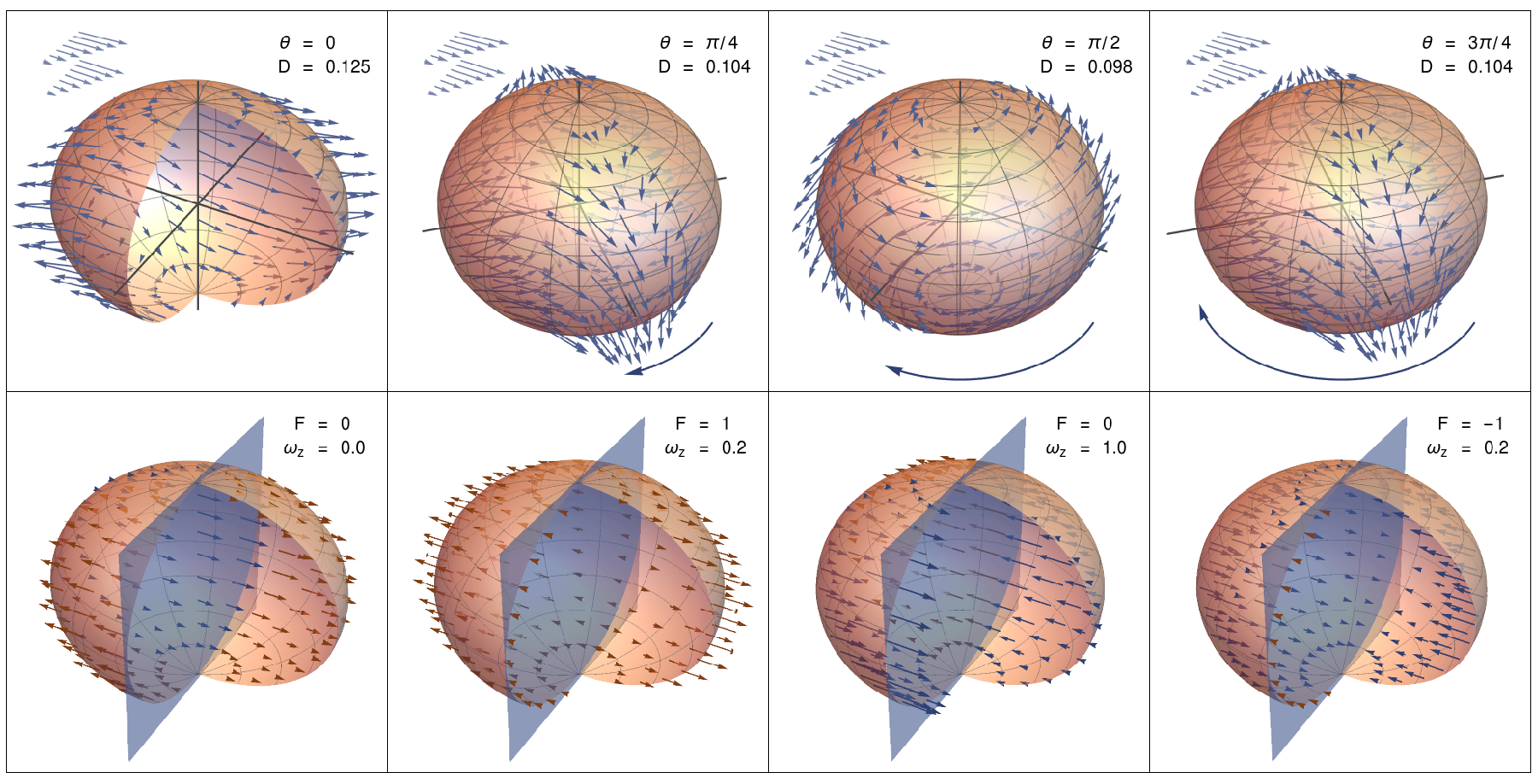}\caption{\label{fig:qualbehav}A sample ellipsoid shown at four characteristic
time points. The ellipsoid is undergoing periodic tumbling with mild
deformation. First row: view from an external reference frame, with
surface forces and flow field. $D=(a_{1}-a_{3})/(a_{1}+a_{3})$ is
the Taylor deformation parameter , $\theta$ is the angle through
which the ellipsoid has rotated. Second row: view from the ellipsoid
reference frame, with components of the surface force acting to push
into (blue) and pull against (red) a sample fragmentation plane. $F$
is the relative fragmentation force with respect to the plane, and
$\omega_{z}$ is the relative angular velocity.}
\end{figure*}

The integrand is thus the signed magnitude of the component of $\mathbf{f}$
acting against $P$, where $s$ indicates whether this component acts
to pull against or push into the plane. This is a significant departure
from our earlier concept of a fragmentation force presented in \cite{ByrneDzulSolomonEtAl2011PhysRevE}
in that $s$ explicitly accounts for the fact that some of the surface
force density may in fact compress against the plane and thus oppose
breakage. A visualization of this phenomenon, as well as the mathematical
constructs used to compute $F$, can be seen in Figure \ref{fig:qualbehav}. 

\subsection{Parameters of the models\label{subsec:parameters}}

The model parameters are described in Table \ref{tbl:parameters}.
The DCM depends upon the initial axes lengths $a_{i}$ of the ellipsoid,
the velocity gradient $\mathbf{L}$, the matrix viscosity $\mu$,
the viscosity ratio $\lambda$, which is the ratio of the droplet
viscosity over the matrix viscosity, and the interfacial tension $\Gamma.$
The FCM depends upon the axes lengths $a_{i}(t)$ at each time point,
the angular velocity of the ellipsoid $\boldsymbol{\omega}(t)$, and
the velocity gradient $\mathbf{L}(t)$, which now also has a dependence
on time due to the rotating frame of reference. 

\begin{table}[h]
\begin{tabular}{ccccc}
\hline 
Symbol & Parameter & Model & Range & Units\tabularnewline
\hline 
$\mathbf{a}(t)$ & axes lengths & D, F & 1-1000$\times10^{-6}$ & m\tabularnewline
$\boldsymbol{\omega}(t)$ & angular velocity & F & 0-100 & 1/s\tabularnewline
$\mathbf{L}(t)$ & velocity gradient & D, F & 0-10 & 1/s\tabularnewline
$\mu$ & matrix viscosity & D, F & $8.95\times10^{-4}$ & Pa s\tabularnewline
$\lambda$ & viscosity ratio & D & 1-1000 & -\tabularnewline
$\Gamma$ & interfacial tension & D & $10^{-9}-10^{-7}$ & N / m\tabularnewline
\hline 
\end{tabular}\caption{\label{tbl:parameters}Model parameters.}
\end{table}

\section{Results\label{sec:Results}}

\subsection{Deformation constituent model\label{subsec:resultsdef}}

\begin{figure}
\centering\includegraphics[width=8.9cm]{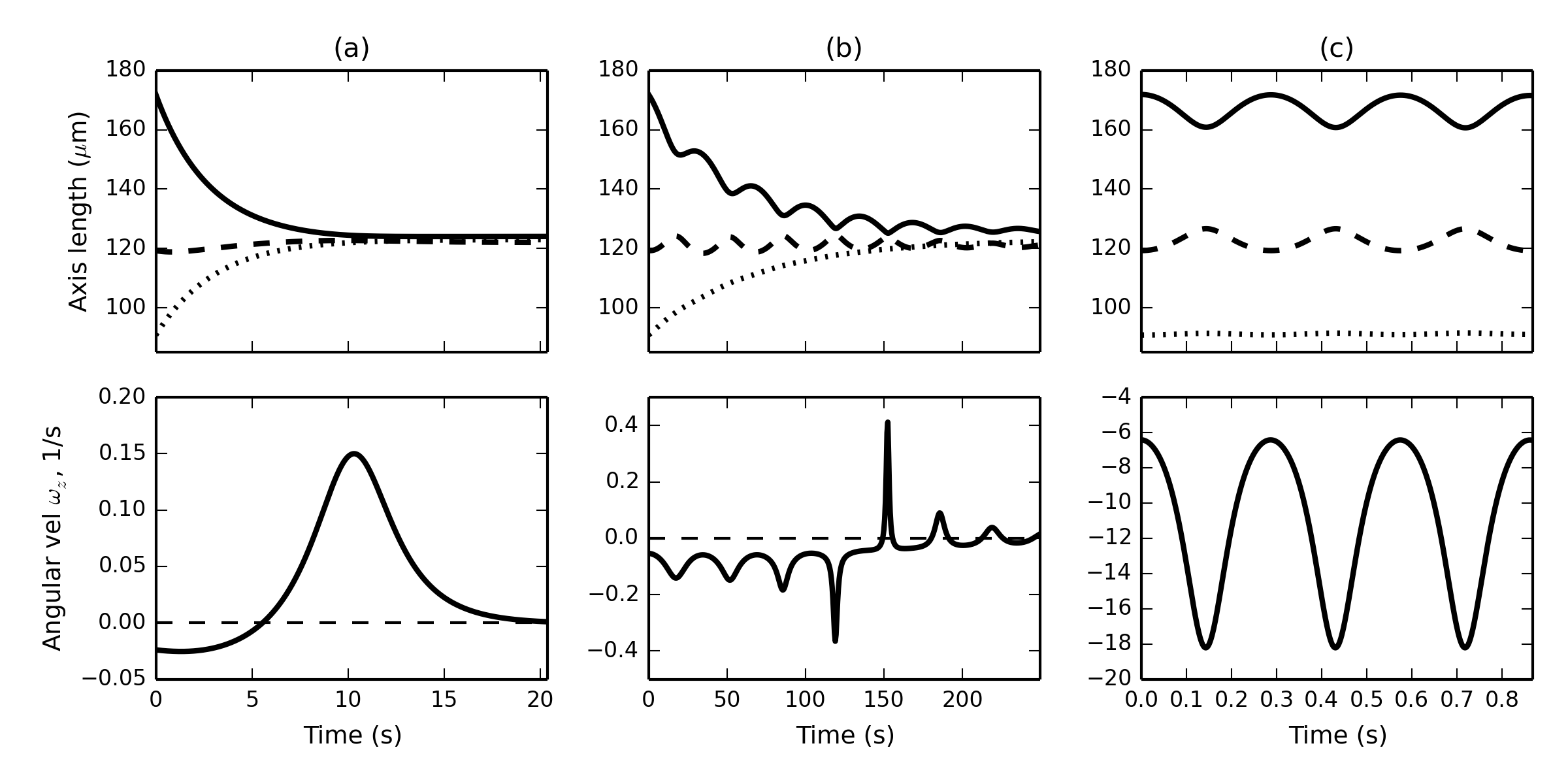}\caption{\label{fig:chardef}Characteristic behaviors of ellipsoids evolving
in the DCM: (a) collapse (Ca$\sim.294$), (b) oscillating collapse
(Ca$\sim13.2$), and (c) periodic tumbling (Ca$\sim3441$). }
\end{figure}

An ellipsoid evolving according to the DCM follows one of three characteristic
behaviors: it can (1) collapse smoothly to a steady-state orientation
and shape (Fig \ref{fig:chardef}a), (2) collapse while oscillating
to a steady-state orientation (Fig \ref{fig:chardef}b), or (3) oscillate
periodically (Fig \ref{fig:chardef}c). The angular velocity of the
oscillating collapse (Fig \ref{fig:chardef}b) can often exhibit a
sudden ``flip'' in which the direction of rotation of the ellipsoid
changes. This occurs when two of the axes lengths are close in magnitude.
This interesting behavior will be the topic of future research, and
in the remainder of the present work we restrict ourselves to the
case of periodic tumbling.

In the oscillatory regime shown in Fig \ref{fig:chardef}c, the behavior
of the deforming ellipsoid approaches that of a solid ellipsoid. In
simple shear defined by $d\mathbf{u}/dy=\dot{\gamma}$, the angle
$\theta(t)$ in the $xy$ plane of a solid ellipsoid is given (\cite{YarinGottliebRoisman1997JFluidMech}
c.f. \cite{Blaser2002ChemEngSci}) by
\begin{equation}
\theta_{solid}(t)=-\arctan\frac{a_{1}}{a_{2}}\tan\left(\frac{2\pi t}{T}\right)
\end{equation}
where 
\begin{equation}
T=\frac{2\pi(a_{1}^{2}+a_{2}^{2})}{a_{1}a_{2}\dot{\gamma}}
\end{equation}
is the period of the rotation. From this we can compute the angular
velocity component $\omega_{z}$ as
\begin{equation}
\omega_{z,solid}=-\frac{2\pi}{T}\frac{a_{1}a_{2}\sec^{2}\left(2\pi t/T\right)}{a_{2}^{2}+a_{1}^{2}\tan^{2}\left(2\pi t/T\right)}.\label{eq:wsolid}
\end{equation}

\begin{figure}[h]
\centering\includegraphics[width=8.9cm]{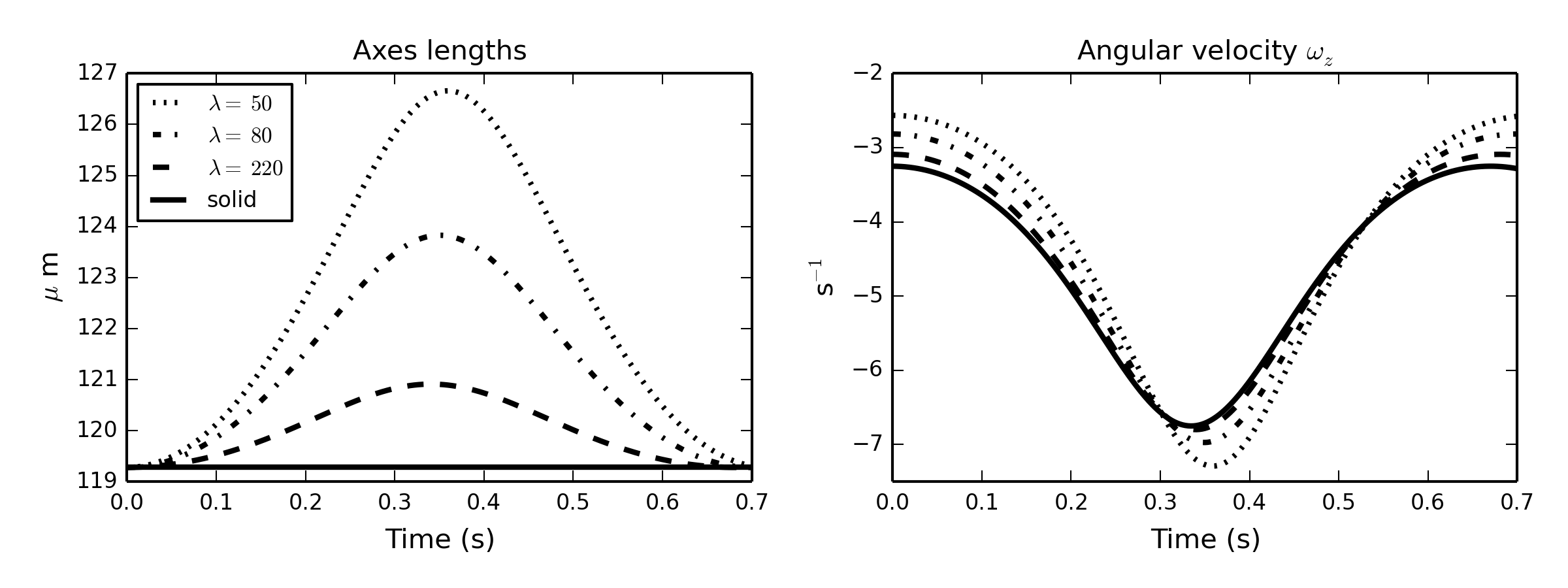}\caption{\label{fig:lamlim}Asymptotic behavior of the DCM as $\lambda\to\infty$
(dashed lines) compared to the behavior of a solid ellipsoid with
angular velocity given by equation (\ref{eq:wsolid}) (solid line).
Left: second axis length ($a_{2})$ over time, right: angular velocity
component $\omega_{z}$ over time. }
\end{figure}

In the limit $\lambda\to\infty$, a fluid droplet becomes a solid,
in which case we expect tthat the axes lengths will become constant
and the angular velocity computed using equation (\ref{eq:wmat})
will approach that given in equation (\ref{eq:wsolid}). This is indeed
what we observe; as the viscosity ratio increases, the axis length
oscillations decrease (Figure \ref{fig:lamlim}a) and the angular
velocity converges to that of a solid ellipsoid (Figure \ref{fig:lamlim}b). 

\subsection{Fragmentation force\label{subsec:resultsfragforce}}

Figure (\ref{fig:qualbehav}) shows the evolution in time of a generic
ellipsoid at four time-points, including the surface force density
as well as the component of this density acting normal to a sample
fragmentation plane. This ellipsoid is undergoing periodic tumbling,
as in Figure (\ref{fig:chardef})c, with mild deformation. In the
first frame, at the initial time, we observe that the surface force
density points both outwards and inwards. This feature is responsible
for the fact that at the third time point, when the angular velocity
is maximized which in turn causes the surface force density magnitudes
to be maximized, we nevertheless observe a net fragmentation force
of 0. The maximum fragmentation force is observed at the second time-point,
when all of the force vectors act against the plane, and the minimum,
which is negative, occurs at the fourth time point, when all of the
surface force vectors push into the plane.

\begin{figure}[h]
\centering\includegraphics[width=8cm]{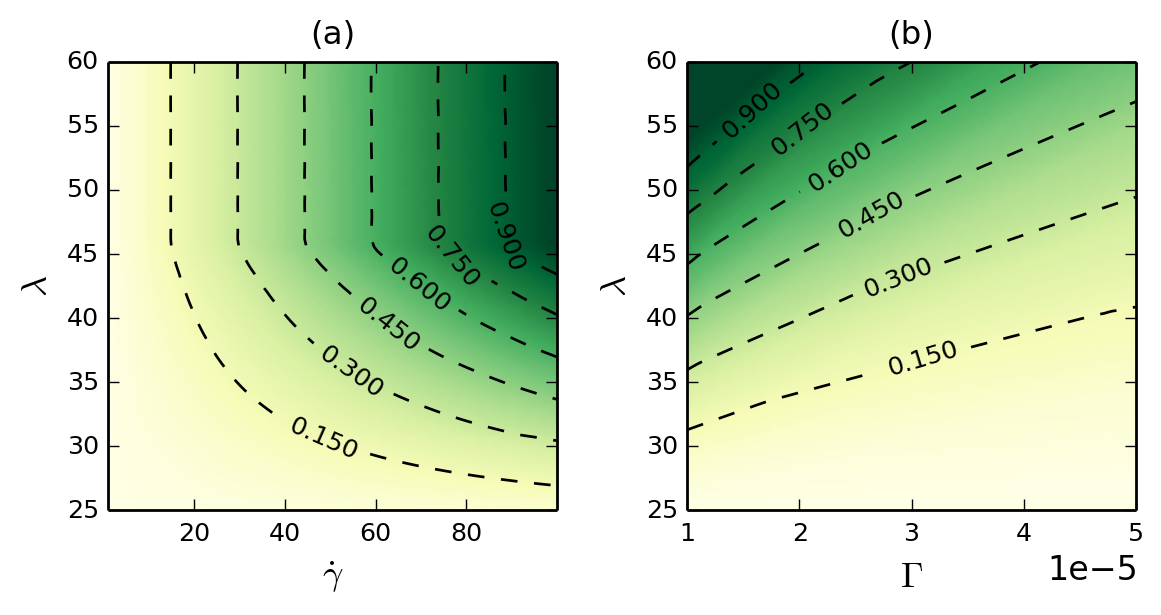}

\caption{\label{fig:pardep}Maximum normalized fragmentation force experienced
by a sample ellipsoid as a function of the shear rate $\dot{\gamma}$
and the viscosity ratio $\lambda$ (a) and the interfacial tension
$\Gamma$ and the viscosity ratio (b). In (a), $\Gamma=4\times10^{-9}$
N/m, and in (b) $\dot{\gamma}=10$ m/s.}
\end{figure}

We examine the behavior of the fragmentation force on a generic ellipsoid
$\mathcal{E}$. We intersect $\mathcal{E}$ with a plane $P$ defined
by the normal $\mathbf{n}_{p}=(1,0,0)$ and interior point $\mathbf{x}_{p}=(0,0,0)$;
i.e., a plane in the $yz$ plane passing through the origin and normal
to the longest major axis of $\mathcal{E}$. We first explore the
dependence of the fragmentation force, equation (\ref{eq:fragforce}),
on the system parameters. We compute the maximum fragmentation force
as a function of the shear rate $\dot{\gamma}$, the viscosity ratio
$\lambda$ (which we vary while holding the matrix viscosity $\mu$
constant, changing only $\mu^{*}$), and the interfacial tension $\Gamma$.
As can be seen in Figure \ref{fig:pardep}(a), the fragmentation force
increases with the shear rate. The shear rate appears directly in
the computation of the surface force density in equation \ref{eq:force},
and indirectly as it affects the angular velocity $\boldsymbol{\omega}$.
At higher shear rates there is a greater dependence of $f_{max}$
on the viscosity ratio, and its dependence on $\lambda$ is non-linear,
changing more for smaller values of $\lambda$ while being constant
at higher values of $\lambda$. The dependence of $f_{max}$ on $\Gamma$
and $\lambda$ is shown in Figure \ref{fig:pardep}(b). Again, $f_{max}$
increases with $\lambda$; in addition, it can be seen to decrease
with $\Gamma$. Neither of these terms appear directly in the force
equation (\ref{eq:force}), and so their influence on $f_{max}$ manifests
through their role in shaping the motion and deformation of the ellipsoid
as in equation \ref{eq:gov}.

\begin{figure}[h]
\centering\includegraphics[width=7cm]{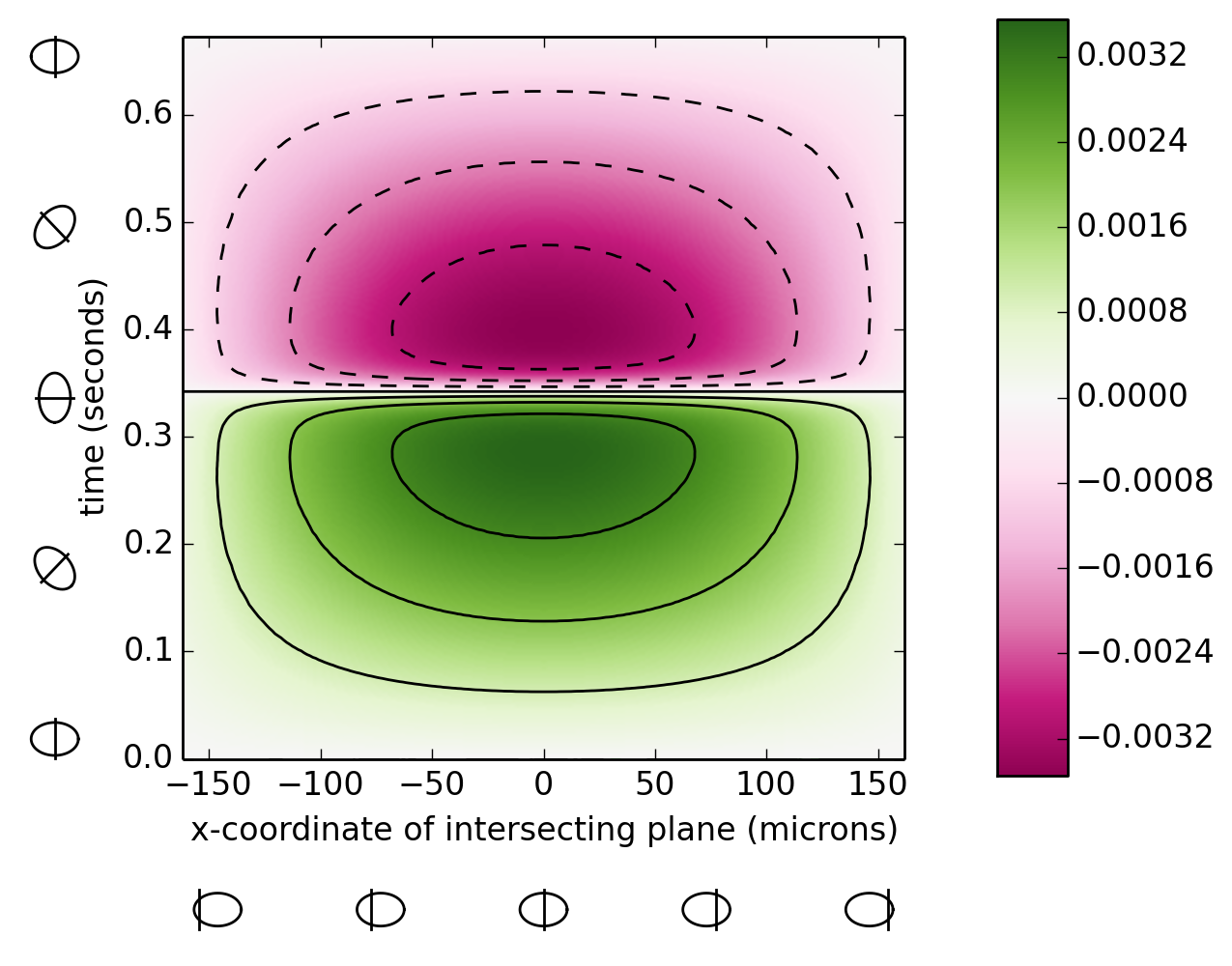}\caption{\label{fig:slidingplane} Fragmentation force experienced by a sample
ellipsoid deforming according to the DCM, as a function of time ($y$
axis) and the position of the intersecting plane ($x$ axis). In this
simulation, $\dot{\gamma}=1$ 1/s, $\lambda=50$, $\mu=8.953\times10^{-4}$
Pa s, and $\Gamma=4.1\times10^{-9}$ N/m. Initial ellipsoid axes are
$\mathbf{a}=(180,160,140)\mu$m.}
\end{figure}

We next consider the fragmentation force as a function of time and
position of the intersecting plane. We construct the ellipsoid as
above, except that now we will slide the intersecting plane along
the $x$ axis. These results are shown in Figure \ref{fig:slidingplane}.
The $x$ axis corresponds to the position of the interior point on
the intersecting plane, so that at a point $x$ on this axis, the
intersecting plane is defined by normal $\mathbf{n}_{p}=(1,0,0)$
and $\mathbf{x}_{p}=(x,0,0)$. The $y$ axis corresponds to time.
The fragmentation force is anti-symmetric about its horizontal center,
which corresponds to the point in time at which the ellipsoid has
rotated through an angle of $\pi/2$. Past this point, the symmetry
of the system results in the surface forces being equal in magnitude
but opposite in sign. As the plane slides along the $x$ axis to the
midpoint, the fragmentation force increases, and then decreases again
as the plane moves from the center to the other end; again due to
the symmetries of the system. 

\section{Conclusion\label{sec:Conclusion}}

We presented a unification of theories used to describe the deformation
of a fluid droplet and the surface forces on a solid ellipsoid in
shear flow. We investigated the qualitative behavior of the forces
on the droplet and its motion, and compared this behavior to the simpler
case of a solid ellipsoid. We introduced the concept of a fragmentation
force, which is the integral of the component of the surface force
density acting against an intersecting plane, and saw how this force
responds to the placement of the intersecting plane and to the deformation
and motion of the droplet. We intend to use the model developed here
to simulate the fragmentation of microbial aggregates; in so doing,
we will extend our previous work \cite{ByrneDzulSolomonEtAl2011PhysRevE}
by permitting the aggregates to deform and by using the more sophisticated
definition of fragmentation force that we have developed here. We
expect the work herein to be applicable more generally to simulations
of colloidal breakage in which it is desirable to preferentially choose
breakage locations, for example in the case of inhomogenous composition.
Code for this work (Python and C) is available on GitHub at \url{https://github.com/MathBioCU/fragforce}.
\begin{acknowledgments}
EPK is supported by the Interdisciplinary Quantitative Biology Program
at the BioFrontiers Institute, University of Colorado Boulder (NSF
IGERT 1144807) and by an NSF GRFP (DGE 1144083). This work was supported
in part by grant NSF-DMS 1225878 to DMB. We would like to thank Charles
Tucker III, Eric Wetzel, and Nancy Jackson for making their code available
to us and for helpful discussions on the behavior of their model,
and Dr. John Younger for helpful discussions on an application of
this work to microbial flocculation. 
\end{acknowledgments}

\bibliographystyle{apsrev4-1}
\bibliography{mathbioCU}

\end{document}